\pdfoutput=1

\documentclass[preprint,12pt,authoryear]{elsarticle}

\usepackage{amssymb}
\usepackage{url}
\usepackage{amsthm}
\usepackage{amsmath}
\usepackage{siunitx}
\usepackage{wrapfig}
\usepackage[export]{adjustbox}
\usepackage{geometry}
\usepackage{array}
\usepackage{tabularx}
\usepackage{caption}
\usepackage{graphicx} 
\usepackage{tabularx}
\usepackage{tablefootnote}
\usepackage{lineno}

\journal{Atmospheric Research}

\begin{document}

\renewcommand{\arraystretch}{1.8}
\newcolumntype{M}{X<{\vspace{4pt}}}

\begin{frontmatter}

\title{Multi-scale assessment of high-resolution reanalysis precipitation fields over Italy}

\author[inst1,inst3,inst2]{Francesco Cavalleri\corref{cor1}}
\ead{francesco.cavalleri@unimi.it}
\author[inst2]{Cristian Lussana}\ead{cristianl@met.no}
\author[inst3]{Francesca Viterbo}\ead{viterbo.francesca@rse-web.it}
\author[inst4]{Michele Brunetti}\ead{m.brunetti@isac.cnr.it}
\author[inst3]{Riccardo Bonanno}\ead{bonanno.riccardo@rse-web.it}
\author[inst1]{Veronica Manara}\ead{veronica.manara@unimi.it}
\author[inst3]{Matteo Lacavalla}\ead{lacavalla.matteo@rse-web.it}
\author[inst3]{Simone Sperati}\ead{sperati.simone@rse-web.it}
\author[inst5]{Mario Raffa}\ead{mario.raffa@cmcc.it}
\author[inst6]{Valerio Capecchi}\ead{capecchi@lamma.toscana.it}
\author[inst7]{Davide Cesari}\ead{dcesari@arpae.it}
\author[inst8,inst7]{Antonio Giordani}\ead{antonio.giordani3@unibo.it}
\author[inst9,inst7]{Ines Maria Luisa Cerenzia}\ead{ines.cerenzia@redrisk.com}
\author[inst1]{Maurizio Maugeri}\ead{maurizio.maugeri@unimi.it}

\cortext[cor1]{Corresponding author}


\affiliation[inst1]{organization={Environmental Science and Policy Department (ESP), University of Milan},
city={Milan},
postcode={20133}, 
state={Italy},
country={Italy}}

\affiliation[inst2]{organization={Division for Climate Services, the Norwegian Meteorological Institute},
city={Oslo},
postcode={0313}, 
country={Norway}}

\affiliation[inst3]{organization={Sustainable Development and Energy Resources Department, Research on Electric Systems (RSE)}, 
city={Milan},
postcode={20134},
country={Italy}}

\affiliation[inst4]{organization={Institute of Atmospheric Sciences and Climate, National Research Council (CNR-ISAC)}, 
city={Bologna},
postcode={40129}, 
country={Italy}}

\affiliation[inst5]{organization={Euro-Mediterranean Center on Climate Change (CMCC)},
city={Caserta},
postcode={81100},
country={Italy}}

\affiliation[inst6]{organization={Environmental Monitoring and Modeling Laboratory (LaMMA)},
city={Sesto Fiorentino (FI)},
postcode={50019}, 
country={Italy}}

\affiliation[inst7]{organization={Regional Agency for Environmental Prevention and Energy of Emilia-Romagna (ARPAE)},
city={Bologna},
postcode={40139}, 
country={Italy}}

\affiliation[inst8]{organization={Department of Physics and Astronomy (DIFA), University of Bologna},
city={Bologna},
postcode={40127}, 
country={Italy}}

\affiliation[inst9]{organization={Risk Engineering + Development (RED)},
city={Pavia},
postcode={27100 }, 
country={Italy}}

\newpage
\begin{abstract}
This study focuses on the validation of high-resolution regional reanalyses to understand their effectiveness in reproducing precipitation patterns over Italy, a climate change hotspot characterized by coastal sea-land interaction and complex orography. Nine reanalysis products were evaluated, with the ECMWF global reanalysis ERA5 serving as a benchmark. These included both European (COSMO-REA6, CERRA) and Italy-specific (BOLAM, MERIDA, MERIDA-HRES, MOLOCH, SPHERA, VHR-REA\_IT) datasets, using different models and parametrizations. 
The inter-comparison involved determining the effective resolution of daily precipitation fields using wavelet techniques and assessing intense precipitation statistics through frequency distributions. In-situ observations and observational gridded datasets were used to independently validate reanalysis precipitation fields.
The capability of reanalyses to depict daily precipitation patterns was assessed, highlighting a maximum radius of precipitation misplacement of about 15 km, with notably lower skills during summer. 
An overall overestimation of precipitation was identified in the reanalysis climatological fields over the Po Valley and the Alps, whereas multiple products showed an underestimation of precipitations across the North-West coast, the Apennines, and Southern Italy.
Finally, a comparison with a time-consistent observational dataset (UniMi/ISAC-CNR) revealed a non-stable deviation from observations in the annual precipitation cumulate of the reanalysis products analyzed. This should be taken into account when interpreting precipitation trends over Italy. 
\end{abstract}


\begin{highlights}
\item Reanalyses enable accurate precipitation reconstruction but require validation.
\item Wavelet analysis reveals higher effective resolution in convection-permitting models.
\item Small-scale precipitation exhibits misplacements up to 15 km, notably in summer.
\item Wet biases in Po Valley/Alps; dry biases on N-W coast, Apennines, Southern Italy.
\item Annual precipitation trends show long-term deviations from observations.
\end{highlights}

\begin{keyword}
reanalysis \sep precipitation \sep validation \sep Italy \sep ERA5 \sep wavelets \sep SEEPS
\end{keyword}

\end{frontmatter}


\section{Introduction}\label{sec:intro}

Precipitation is a fundamental component and the main driver of Earth's water cycle. Understanding and accurately reproducing precipitation fields presents significant challenges, as precipitation phenomena are intermittent and highly variable in space \citep{sun_2018, Schleiss_2011}. In-situ weather observations provide an accurate estimate of precipitation, but only at a limited number of locations. These observations are sometimes neither dense nor temporally extensive enough to fully represent the precipitation processes of interest. Observational gridded datasets, instead, are derived from statistical interpolation of station observations, to better depict the spatial variability of precipitation. However, they may suffer from temporal and spatial incompleteness \citep{prein_2016}. Conversely, atmospheric reanalyses, that combine the use of atmospheric dynamical models with the assimilation of multiple sources of observations \citep{Kalnay_2024}, provide temporally and spatially coherent datasets, with physical relationships among variables. Reanalysis products enable the reconstruction of past climate conditions both for scientific research and practical applications \citep{simmons_2017}, although errors may arise from potential inhomogeneities of assimilated observational data, limitations in modelling physical processes, and challenges in representing complex topography \citep{rapaic_2015}.

In recent years, the growing significance of reanalysis datasets and their multiple applications across various disciplines have led to an increased availability of these products, at different resolutions and over different spatial domains. These products range from global reanalyses, which cover the entire globe at resolutions of approximately 200 km to 30 km, to regional products that focus on specific areas with higher resolutions (10 km to 5 km). Some regional products also employ non-hydrostatic models, allowing the explicit representation of convection at scales below 4 km. The added value of regional reanalyses in reconstructing precipitation phenomena, relative to global reanalyses, has been highlighted in numerous studies \citep{kaiserweiss_2019,Lucas‐Picher_2017}.

In this study, nine different reanalysis products, both global and regional, are selected to assess their performance over Italy for the period from 1995 to 2019. This timeframe represents the longest common period among the analyzed reanalyses.
ERA5 \citep{hersbach_2020_era5} is selected as the global benchmark, while regional reanalyses, including both pan-European and Italy-specific products, are chosen to provide a diverse set of models, parameterizations, and assimilated datasets. This selection aims to enable a wide evaluation of reanalyses from different model designs. Additionally, most of these regional products have been recently released, necessitating an inter-comparison and independent validation beyond the assessments conducted by their developers. While surface air temperature fields from some of these models have undergone such validation \citep{cavalleri_2024}, this work aims to address precipitation fields.

Numerous assessments of precipitation fields from the global reanalysis ERA5 have been conducted in recent years, covering global scales \citep{lavers_etal_2022,lavers_etal_2023} as well as specific regions worldwide \citep{Beck_2019,Cardoso_2024,Crossett_2020,Dollan_2024,Gheysari_2024,Jiang_2024,Kislov_2022,Tarek_2020}. However, only a limited number of studies have specifically focused on Europe \citep{Alexopoulos_2023,Gomiscebolla_2023,Hassler_2021} and Italy \citep{adinolfi_2023,lussana_cavalleri_2024}. In particular, \cite{bandhauer_2022} evaluated ERA5 and E-OBS performances across three European subregions (the Alps, the Carpathians, and Fennoscandia) using high-quality regional datasets derived from dense rain-gauge data as references, such as the Alpine Precipitation Gridded Dataset \cite[APGD,][]{isotta2014climate} for the Alps. In this work, the authors noted that the spatial resolution of ERA5 appears generally coarser than its formal grid spacing, indicating a need for further investigation in this aspect.
Notably, \cite{isotta_2015} compare the global reanalysis ERA-Interim \citep{dee_erainterim_2011}, some different regional reanalyses and downscaling datasets, and station-based interpolation datasets against a reference gridded observational dataset (APGD). Many of the studies cited so far have provided useful insights for developing evaluation methodologies that are applicable to the most recent reanalysis products.

In this context, the first objective of this work is to conduct an intercomparison of daily precipitation fields derived from various reanalyses to quantify their effective resolutions, which quantifies the characteristic length scale of the spatial filter employed in the convolution of a continuous precipitation field onto the gridded domain of the model, as elucidated in Section \ref{sec:methods_effRes}.
Providing an accurate estimate for the effective resolution of a reanalysis is important, and constitutes one of the innovative parts of this work. Indeed, this value is frequently only approximately known to both model users and developers.
Then, the study aims to analyze the frequency distributions of daily precipitation exceeding specific thresholds to explore the relationship between the effective spatial resolution and the representation of intense precipitation events. Once the effective resolution of the reanalyses has been assessed, the accuracy of the reanalysis products against observational data is verified. In addition to that, reanalysis daily precipitation fields are compared against a dense network of weather stations (Section \ref{sec:dewetra}), to quantify the uncertainty in rainfall positioning and to explore the skills and limits in predictability of regional reanalyses. Moreover, a spatial and seasonal assessment of normal precipitation values from different products is performed by comparing monthly climatological reanalysis fields with the corresponding climatologies from the observational gridded dataset UniMi/ISAC-CNR (Section \ref{sec:unimi}). The same dataset is used to evaluate the skill of reanalyses in capturing long-term precipitation trends by assessing the temporal consistency of reanalysis precipitation annual amounts.

The paper is organized as follows: Section~\ref{sec:data} provides a description of the reanalysis and the observational datasets used in this study. Section~\ref{sec:methods} outlines the methodology adopted for evaluating the products under investigation. Subsequently, Section~\ref{sec:results} presents and discusses the results of the analyses. Finally, in Section~\ref{sec:conclusions}, the main findings are summarized and their implications are discussed.

\section{Data}\label{sec:data}

\subsection{Reanalyses}\label{sec:data_rea}

In this study, different types of reanalyses were considered. The global reference is ERA5 \citep{hersbach_2020_era5}, the latest reanalysis developed by the \textit{European Centre for Medium-range Weather Forecasts} (ECMWF). The parameterized-convection products analyzed include the reanalysis \textit{Consortium for Small-scale Modeling REAnalysis at 6 km} (COSMO-REA6, \cite{bollmeyer_cosmorea6_2015}, $\approx$ 6 km grid spacing), the reanalysis \textit{Copernicus European Regional ReAnalysis} (CERRA, \cite{schimanke_2021_cerra}, $\approx$ 5.5 km grid spacing), a hindcast which uses the \textit{Bologna Limited Area Model} (BOLAM, \cite{vannucchi_2021_bolam}, $\approx$ 7 km grid spacing), and the dynamical downscaling of ERA5 \textit{MEteorological Reanalysis Italian DAtaset} (MERIDA, \cite{bonanno_2019_merida}, $\approx$ 7 km grid spacing).
Convection-permitting products, capable of explicitly resolving convection \citep{prein_2015}, were also included in the study. These consist of a hindcast which uses the \textit{MOdello LOCale in Hybrid coordinates} (MOLOCH, \cite{capecchi_2023_moloch}, $\approx$ 2 km grid spacing), and three dynamical downscalings of ERA5: the \textit{MEteorological Reanalysis Italian DAtaset High-RESolution} (MERIDA-HRES, \cite{viterbo_2024}, $\approx$ 4 km grid spacing), the \textit{Special Project: High rEsolution ReAnalysis over Italy} (SPHERA, \cite{cerenzia_2022,giordani_2023_sphera}, $\approx$ 2 km grid spacing), and the \textit{Very High Resolution Dynamical Downscaling of ERA5 Reanalysis over Italy} (VHR-REA\_IT, \cite{raffa_2021_vhr,reder_2022}, $\approx$ 2 km grid spacing).

Each of these reanalysis products gets the initial and boundary conditions from ERA5, except for COSMO-REA6, which employs boundary and initial conditions from the predecessor of ERA5, ERA-Interim \citep{dee_erainterim_2011}, and MOLOCH, which is nested into BOLAM.
In terms of observation assimilation, CERRA operates with a 3D-Var data assimilation scheme, COSMO-REA6 and SPHERA utilize an observational nudging scheme, MERIDA and MERIDA-HRES employ a spectral observational nudging scheme, while MOLOCH, BOLAM, and VHR-REA\_IT do not assimilate data. A summary of the different characteristics of the analyzed reanalysis products is provided in Table~\ref{tab:nwp_comparison}.

\subsection{Observational datasets}\label{sec:data_obs}

In this study, two observational datasets are used as references for validation to provide insights from different viewpoints: a weather stations dataset (Section \ref{sec:dewetra}) spanning from 2016 to 2020, and a gridded observational dataset (Section \ref{sec:unimi}) spanning from 1995 to 2019.

\subsubsection{Weather stations}\label{sec:dewetra}
The weather station data utilized in this study were obtained from the observational networks operated by the Regional Agencies for Environmental Protection (ARPA), evenly distributed throughout the Italian national territory. These stations measure hourly precipitation and send data in real-time to the operational platform used by the Italian Civil Protection Department to streamline operational activities \citep{dewetra}. Although measurements are available dating back to 2002, it was only from 2016 onwards that the spatial coverage became homogeneous enough for validation purposes. A 5-year period from 2016 to 2020 was chosen for the comparison with the reanalysis ERA5, MERIDA, MERIDA-HRES, CERRA, VHR-REA\_IT, while a 4-year period from 2016 to 2019 was used for reanalyses missing the year 2020 (COSMO-REA6, BOLAM, MOLOCH) or for which this year was not available during the proceedings of this analysis (SPHERA).

The same techniques as those described in \cite{bonanno_2019_merida} were used to quality-check these stations. An essential aspect of this validation involved discarding precipitation data associated with temperatures below 2°C to mitigate the possible undercatch caused by snow \citep{Barry_1978}.
Furthermore, a robust subset of stations was selected, providing data for at least 25\% of the days during the 2016-2020 period, also ensuring the condition of at least 25\% data in each season. From an initial pool of 2870 unique weather stations, the quality check left 1622 stations for annual analysis. For seasonal analyses, the subsets consist of 1625, 2052, 2159, and 2141 stations for winter, spring, summer, and autumn, respectively.

\subsubsection{UniMi/ISAC-CNR gridded dataset}\label{sec:unimi}
Monthly precipitation maps for Italy spanning from 1995 to 2019 were generated. These maps utilized a dense weather station dataset situated across Italy and its northern neighbouring regions. The dataset, which includes a subset of the station data detailed in Section \ref{sec:dewetra}, underwent homogenization and quality verification. Furthermore, it encompasses data from the Italian Air Forces, ENAV (the Italian Air Traffic Control and Assistance Company), the Swiss Meteorological Service (MeteoSwiss), and the HISTALP dataset \citep{histalp_2007}. Specifically, a digital elevation model with a resolution of 30 arc-seconds was employed to produce precipitation fields by interpolating measurements onto its grid. This process is based on the anomaly method \citep{Mitchel_etal_2005,Brunetti_etal_2012,crespi_2021}. The methodology focuses on independently reconstructing monthly climatologies (mean values estimated over a specific reference period) and anomalies (the deviations to the climatologies). Climatologies exhibit pronounced spatial gradients, necessitating a substantial number of weather stations (even if available for a brief period) for being accurately captured, along with an interpolation technique that leverages the correlation between climate variables and geographical parameters. Anomalies, influenced by climate change and variability, demonstrate higher spatial coherence and can be captured by a limited number of stations using a simpler interpolation technique, albeit with the prerequisite of data homogenization. Ultimately, monthly precipitation fields are derived by overlaying the reconstructed climatologies and anomalies. The climatologies are obtained over a 30-arc-second resolution grid, using a local weighted linear regression of precipitation normals from neighboring stations (from a dataset of 6134 stations) against elevation, with weights based on the similarity (horizontal and vertical distance, slope steepness and orientation, distance from the sea) between the stations and the grid cell \citep{crespi_2018}. Anomaly records are calculated over the same grid, starting from a dataset comprising a number of series between 1000 and 3000 over the 1995-2019 period. These records are derived as weighted averages of anomalies from stations surrounding each grid point. These weights are a combination of radial and vertical weighting functions, supplemented by an angular weight to consider anisotropy in the distribution of stations around the grid point \citep{González-Hidalgo_etal_2011}. The robustness of the resulting UniMi/ISAC-CNR dataset has been furtherly corroborated in \cite{lussana_cavalleri_2024} by comparing it with the observational dataset LAPrec \cite{Isotta_etal_2024} in overlapping regions.

\section{Evaluation strategy}\label{sec:methods}

\subsection{Intercomparison among reanalyses}\label{sec:methods_int}

The primary aim of the intercomparison among different reanalyses is to understand what "effective resolution" means and to study how it varies across different reanalysis products (Section \ref{sec:methods_effRes}). The strategy involves designing a method to classify a reanalysis precipitation dataset as either global or regional, based solely on the analysis of its daily precipitation fields.
Additionally, the goal is to determine if it is possible to distinguish between convection-permitting models and those with parameterized convection, without prior knowledge of the model design. The next step is to revisit the reanalysis production cycle to understand the sources of uncertainty in reanalysis data and how limitations in predictability emerge (Section \ref{sec:predictability}).
Moreover, the relationships between the effective resolution of precipitation fields and the precipitation phenomena reconstructed by the models are explored, with a particular emphasis on intense precipitation events (Section \ref{sec:met_distr}).

\begin{figure}[ht!]
    \centering
\includegraphics[width=1\textwidth,left]{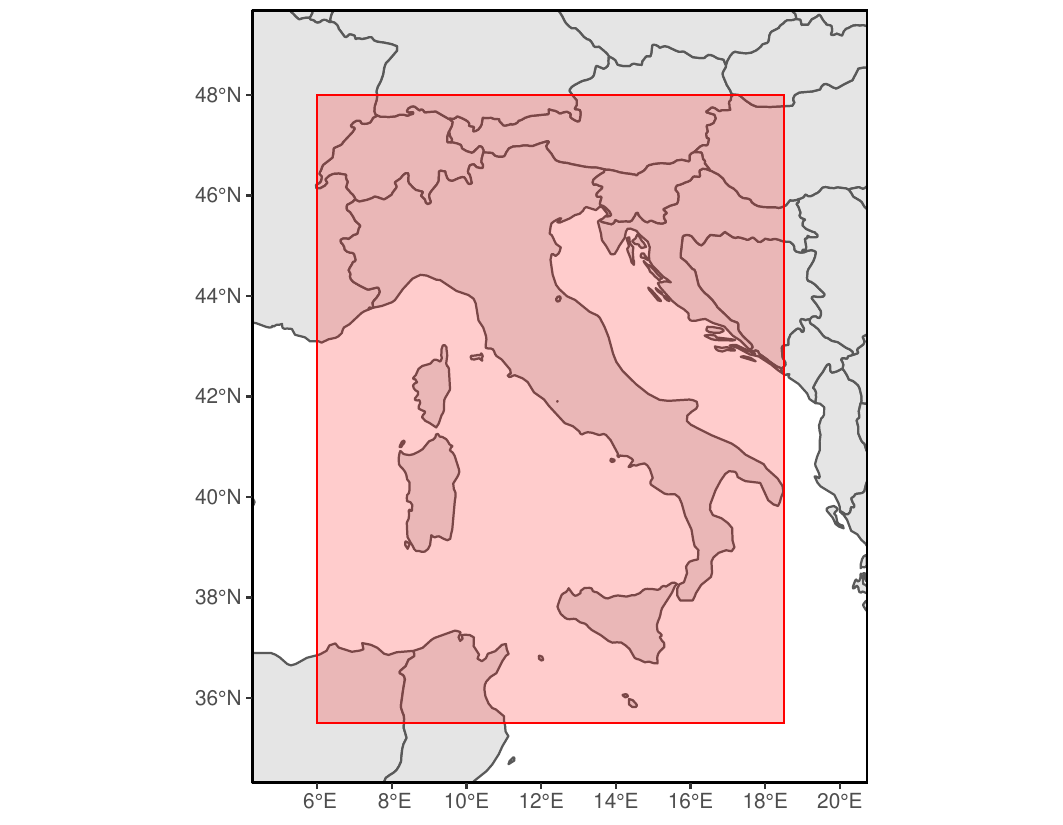}
    \caption{\textbf{Spatial Domain for Inter-Comparison} The spatial domain inside which inter-comparisons among reanalyses were carried out (highlighted in red).}
    \label{fig:dyadicdomain}
\end{figure}

These inter-comparisons were conducted over the domain depicted in Figure \ref{fig:dyadicdomain} that encloses both land and sea points of the Italian region. The results of the inter-comparison are presented in Section \ref{sec:inter-comp}.

\subsubsection{Scale-separation diagnostics based on wavelet decomposition}\label{sec:methods_effRes}

When precipitation data are provided on a regular grid, it is essential to state explicitly how the cell value is interpreted. From a mathematical point of view, the cell value represents the convolution of the continuous precipitation field by a spatial filter (or kernel), as described in Eq.(1) of \cite{Frehlich_2011}.
A widely used approximation for this convolution process assumes that the value for each grid cell represents the mean precipitation within that cell, effectively treating the spatial filter as coextensive with the grid box itself. This approximation has been applied to ERA5 precipitation data by \cite{lavers_etal_2022}, who assert that each ERA5 grid value averages precipitation over a $0.25^\circ \times 0.25^\circ$ grid box, approximating areas of about 1000 km$^2$. 
Other interpretations are possible, for instance, the ERA5 User Guide (\url{https://confluence.ecmwf.int/display/CKB/ERA5%3A+What+is+the+spatial+reference}, last accessed 05/04/2024) proposes viewing the data as representing point values at regular intervals. The point of view adopted in this work is that there is a critical distinction between field resolution and grid spacing \citep{grasso2000_bams}. The grid spacing of a field inherently sets a minimum threshold for the characteristic length scale of the spatial filter employed in its convolution. The effective resolution of a precipitation field aims to quantify this characteristic length scale.

The method used in this work to determine the effective resolution of the reanalyses is a scale-separation diagnostic based on wavelet decomposition (\cite{briggs_1997,casati_2004,casati_2007,jung_2008scale,casati_2010}). The daily scale is chosen because it enables focusing on the reconstruction of precipitation phenomena down to the meso-$\beta$ scale \citep{thunis_scales_1996}, ranging from an upper boundary of 200 km to a lower boundary of 20 km, and potentially extends to even finer scales, down to the meso-$\gamma$ scale (i.e., from 20 km to 2 km). Indeed, daily precipitation is defined as the total precipitation over a fixed 24-hour window, allowing it to capture precipitation events of significantly shorter duration, such as a single thunderstorm lasting only a few hours. The effective resolution of regional reanalyses should reasonably align with the meso-$\beta$ or meso-$\gamma$ scales. This expectation is based on the grid spacings of regional models, which span only a few kilometers, and on the hourly aggregation of total precipitation, representing the minimum temporal aggregation typically utilized.

The technique used in this work refers in particular to the 2D Haar discrete wavelet transforms used by \cite{casati_2023_sbe}. Each daily precipitation field is modelled as the cumulative result of multiple components, where each component delineates the contribution from a distinct spatial scale and is mapped onto a regular grid. The spacing of each grid mirrors the respective spatial scale. Each component is obtained by some coefficients that multiply the wavelet bases, which are an orthonormal basis for the space of the discrete precipitation signal. The bases are immutable across spatial scales; they undergo spatial translation and are scaled up or down to align with grids of varying spatial scales. Conversely, wavelet coefficients are determined to ensure the aggregated spatial components reconstruct the original precipitation field. Consequently, if a daily precipitation field comprises predominantly large-scale precipitation events, such as precipitation from stratiform clouds in winter, the wavelet coefficients for smaller scales will be nearly zero. In contrast, if the precipitation field is characterized by numerous small-scale events, like convective episodes, the wavelet coefficients for these smaller scales will markedly deviate from zero.
Reanalyses with varying effective resolutions will exhibit systematic discrepancies in the reconstruction of daily precipitation fields, particularly in stratiform and convective events in the meso-$\beta$ and -$\gamma$ scales. This is due to the effective resolution setting a minimum threshold for the spatial scale at which wavelet coefficients substantially deviate from zero. 
The core concept of wavelet-based decomposition lies in the averaging of statistics of wavelet coefficients over several years, thereby highlighting the characteristics of each reanalysis associated with its specific effective resolution. Through comparison of these statistics, deeper insights can be gained into the relative capabilities of different reanalyses in reconstructing daily precipitation fields.

The procedure is explained in detail by \cite{lussana_2024}, where the energy of a daily precipitation field at a specific spatial scale is defined as being proportional to the variance of the wavelet coefficients associated with that scale, expressed in units of $\mathrm{mm}^2$. To facilitate the inter-comparison of energies between reanalyses, the energies can be presented as percentages of the total energy.

In practical terms, the precipitation field is described at $n$ different spatial scales on a dyadic grid (i.e. a $2^n \times 2^n$ grid, with $n$ a positive integer). In this work, the more detailed dyadic grid has $512 \times 512$ points covering the domain depicted in Figure \ref{fig:dyadicdomain}. Its grid spacing in both axes is approximately $\approx 0.02^\circ$ degrees of latitude and longitude, corresponding to $\approx 2$ km, which is the lower bound of the meso-$\gamma$ scale.
The precipitation fields from each reanalysis are bilinearly interpolated onto this dyadic grid. The bilinear interpolation does not impact the effective resolution of the precipitation fields; rather, the grid spacing is altered while preserving the spatial features within the precipitation fields. The percentage energy at various scales for each daily field was obtained, and then the daily spectra were averaged over the common period shared by all products (1995-2019).
These averaged spectra can be interpreted as the fractions of daily and sub-daily precipitation events occurring at different spatial scales.


\subsubsection{Predictability limitations in global and regional reanalyses}\label{sec:predictability}

The production of a reanalysis involves a two-step iterative process. Firstly, the initial conditions are created through data assimilation, so that the model state is adjusted to better match observational data. The subsequent step involves the production of a forecast, where the model state is advanced through a specified lead time, typically ranging from 6 to 12 hours, depending on the specific design of the reanalysis. This two-step process is iterative, so that observational data are constantly integrated into the model, continually updating and refining the atmospheric conditions with observations over time \citep{Kalnay_2024}.
The nature of predictability limitations at convection-permitting scales for quantitative precipitation forecasting is well known and has been discussed in several studies. In the following, reference will be made to the research presented by \cite{hohenegger_predictability_2007}. Interestingly, this study focused on temperature and geopotential height, variables generally considered more predictable than precipitation.
The limitations of predictability are especially pronounced in terms of the error growth rate and the corresponding error saturation, defined as the lead time at which a specific variable reaches its maximum uncertainty level. Notably, error saturation is observed to occur approximately 10 times earlier in local models compared to global models. 

In local models, the typical time scale for perturbation doubling, or the period over which uncertainty spreads across the field, is approximately four hours. At cloud-resolving scales, the tangent-linear approximation for error growth ceases to be applicable after relatively short integration periods, approximately 1.5 hours. This highlights the pronounced non-linearity in short-term forecasts at these scales. To put this into perspective, a 10-day forecast at synoptic scales can be likened to a cloud-resolving simulation with a lead time of about 7 hours in terms of the rapid escalation of forecast uncertainty.
It is noteworthy that the typical time scales associated with the spread of uncertainties at convection-permitting scales closely align with the integration windows employed in numerical model forecasts within the reanalysis production cycle. Indeed, regional reanalyses aim to reconstruct smaller-scale precipitation patterns compared to their global counterparts. At these finer scales, the inherently chaotic nature of the atmosphere leads to a more rapid increase in uncertainty, impacting both the amplitude and the spatial accuracy of simulated precipitation events. Consequently, with the enhancement of the effective resolution of reanalyses, there is also an anticipation of an escalation in the uncertainty surrounding the magnitude and location of these events.

An inter-comparison among different reanalyses will be conducted to study how the representation of uncertainty changes with the increase in effective resolution, as defined in Section \ref{sec:methods_effRes}. The impact of effective resolution on the accuracy and precision of reanalyses against observations will be examined, with a more specific focus on misplacement errors (see Section \ref{sec:met_seeps}). These inaccuracies, where predicted rainfall events occur at slightly incorrect locations, are crucial because even a minor spatial shift can cause significantly different hydrological responses, potentially leading to flooding in an unintended basin or affecting another watershed. Accurately predicting the location of intense rainfall is essential for anticipating and mitigating the impacts on specific river basins \citep{Paschalis2014On}.

\subsubsection{Frequency distributions of daily precipitation}\label{sec:met_distr}

When a reanalysis can reconstruct daily precipitation fields with an effective resolution at the meso-$\gamma$ scale, it holds the potential to realistically simulate phenomena such as thunderstorms or groups of thunderstorms. These phenomena are characterized by high precipitation values observed over short timescales, ranging from a few hours to one day. Conversely, reanalyses with an effective resolution exceeding the meso-$\beta$ scale tend to poorly reconstruct thunderstorms \citep{Slivinski2019Towards}, often resulting in the underestimation of peak precipitation values. This suggests that a coarser effective resolution model, compared to a finer effective resolution one, tends to smooth the precipitation signal.

To assess this aspect, frequency distributions of daily precipitation fields were computed for the period 1995-2019, using thresholds ranging from 1 to 100mm and 1 mm bins. This technique helps to understand the link between effective resolution and the frequency of simulated peak precipitation values \citep{frei_2003,isotta_2015,napoli_2023}. First, the number of days exceeding the thresholds was calculated and summed over all the grid points of the domain. Then, distributions were normalized by the total number of days (which remains the same for each product) and the different number of grid points (which depends on the grid spacing of each product). 

\subsection{Validation against observations}\label{sec:methods_verifobs}

The comparison between observational data and reanalyses serves various purposes in this study. Firstly, it enables the evaluation of the ability of reanalyses to accurately reconstruct daily precipitation fields, particularly focusing on uncertainties due to spatial misplacements of precipitation events (Section \ref{sec:met_seeps}). Secondly, it aids in assessing the skills of reanalyses in accurately reproducing the Italian precipitation climatology, and enhances the understanding of the reliability of reanalyses in representing climatic trends (Section \ref{sec:met_gridd}). 

\subsubsection{Comparison against in-situ observations}\label{sec:met_seeps}

To compare daily precipitation fields against weather station data (see Section  \ref{sec:dewetra} for details about observations) each grid point of each reanalysis which is the closest to a weather station location was selected. This approach is frequently used for precipitation verification, as also used in ECMWF operational verifications \citep{lavers_etal_2022}. The SEEPS score (Stable Equitable Error in Probability Space) was calculated to evaluate the performance of reanalysis products in distinguishing daily precipitation values across the categories of 'dry', 'light precipitation', and 'heavy precipitation' \citep{haiden_2012,rodwell_2010}.
Precipitation has been evaluated considering a categorical approach, which is considered more suitable for validating precipitation data \citep{Rivera_2018}, even if many other studies consider precipitation as a continuous variable \citep{rossa2008overview}. This preference arises because both reference observations and reanalyses are subject to uncertainties that, in the case of precipitation, adhere to a multiplicative error model \citep{tian_2013}, as opposed to other meteorological variables such as temperature, which follow an additive error model. Consequently, the magnitude of uncertainty in precipitation measurements increases with the value of the observed or estimated precipitation, affecting both the reference data and the reanalyses under verification. Thus, a categorical approach to verification offers a more robust and reliable framework for statistical analysis, accommodating the specific error characteristics inherent in precipitation data.
For SEEPS calculation, 'dry' days were defined as those with precipitation accumulation less than 1 mm. The probability $ p_1 $, defined as the fraction of dry days during the reference period, was computed for each station. Stations with $ p_1 < 0.1 $ or $ p_1 > 0.85 $ were excluded to mitigate sampling uncertainties in extreme climates. Then, categories of 'light' and 'heavy' precipitation were based for each station on its climatology during the reference period, so that light precipitation occurs twice as frequently as heavy precipitation on average, as suggested by \cite{haiden_2012}. The reference period is 2016-2020 or 2016-2019 depending on reanalysis data availability as noted in Section \ref{sec:dewetra}.

A 3x3 contingency table (Equation \ref{eq:SEEPS}, left matrix) is populated with the fraction of days in each category, and the SEEPS is computed as the scalar product of this table with the scoring matrix $S$ (Equation \ref{eq:SEEPS}, right matrix) as described by \cite{haiden_2012}. In this study, a skill score (SS) defined as $1 - \text{SEEPS}$ is used to provide a skill assessment ranging from a value of $0$ for no skill to the best value of $1$ for perfect skill. 

\begin{equation}
\mbox{\fontsize{9.5}{21.6}\selectfont\( %
\hspace{-1cm}
SS = 1 \; - 
\quad
\bordermatrix{     
                & \text{dry obs} & \text{light obs} & \text{heavy obs} \cr
\text{dry rean} & ...            & ...              &  ...             \cr
\text{light rean} & ...            & ...              &  ...             \cr
\text{heavy rean} & ...            & ...              &  ...             \cr
}
\;
\cdot 1/2
\bordermatrix{     
     &         &         &        &       \cr
     & 0       &  \frac{1}{1-p_1} &  \frac{4}{1-p_1}    \cr
     & \frac{1}{p_1}  & 0         & \frac{3}{1-p_1}       \cr
     & \frac{1}{p_1} + \frac{3}{2+p_1}  & \frac{3}{2+p_1}  & 0 &    \cr
}
\label{eq:SEEPS}
\)}
\end{equation}

\vspace{1cm}
SS was calculated daily for every station that met the quality criteria (see Section \ref{sec:dewetra}) on that day, then averaged over the reference period and for each season within it. The results of the validation against in-situ observations are presented in Section \ref{res:daily}.

\subsubsection{Comparison against observational gridded dataset}\label{sec:met_gridd}

Monthly precipitation averages were calculated to obtain the climatology for the period  1995-2019. Since the effective resolution of each reanalysis and the misplacement errors were already assessed in the previous analyses, the comparison was performed at the coarsest resolution among all the reanalysis products (i.e. ERA5). The upscale was performed using a conservative remapping technique, averaging all the grid points falling within an ERA5 grid cell. The same methodology was applied to the 30 arc-second resolution gridded observational dataset UniMi/ISAC-CNR (Section \ref{sec:unimi}). 

A comparison between reanalysis and observational fields was conducted to assess the differences in the spatial distribution and the average seasonality of precipitation, utilizing the percentage bias (Equation \ref{eq:percbias}). This indicator was selected because a relative indicator is more effective in depicting deviations independently of diverse precipitation quantities at different locations and across the year \citep{tian_2013}.

\begin{equation}
    \%bias = \frac{tp_{rean} - tp_{obs}}{tp_{obs}}
    \label{eq:percbias}
\end{equation}

Finally, the yearly difference between the reanalysis and the observations was calculated, resulting in a time series of annual deviations from observations of total precipitation over Italy for each reanalysis, spanning from 1995 to 2019. If reanalyses accurately reproduce the long-term precipitation trends, these series should show a flat trend over the analyzed period. This approach aims to determine the extent to which changes in the statistical properties of reanalysis precipitation quantities correspond to actual shifts in the underlying precipitation climatology \citep{lussana_cavalleri_2024}.
The results of the validation against the UniMi/ISAC-CNR gridded dataset are presented in Section \ref{sec:res_clim} and Section \ref{sec:res_trend}.

\section{Results and Discussion}\label{sec:results}

\subsection{Intercomparison: effective resolution and precipitation intensity}\label{sec:inter-comp}

\begin{figure}[h!]
    \centering
    \includegraphics[width=1.1\textwidth]{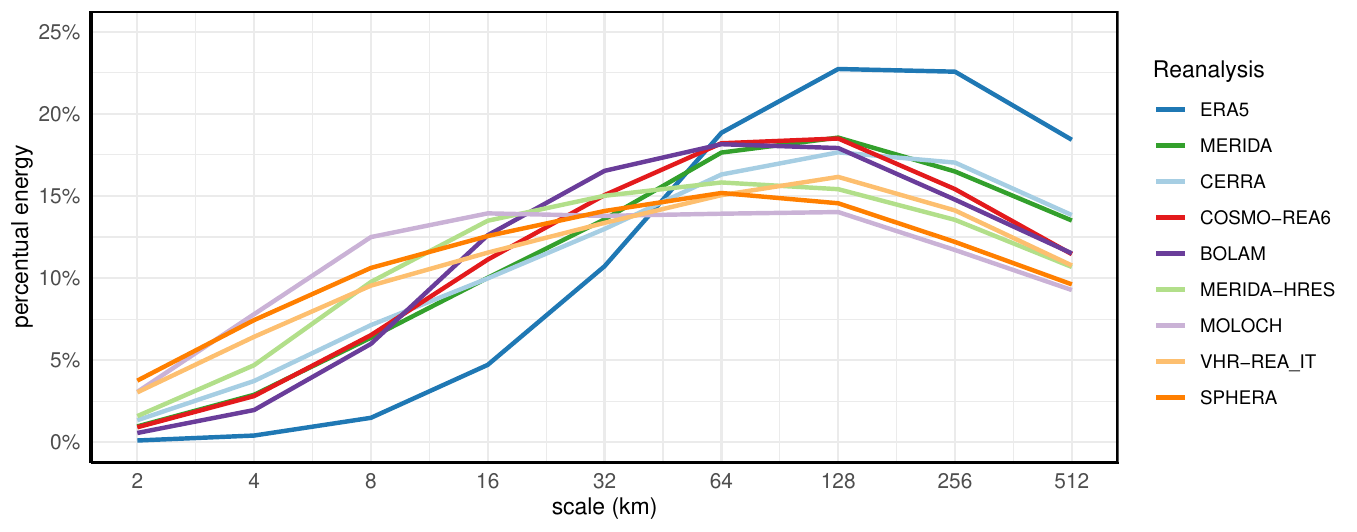}
    \caption{\textbf{Percentage Energy Spectra from Wavelet Decomposition.} \\The percentual energy (y-axis) across different spatial scales (x-axis) obtained from the wavelet decomposition of daily precipitation fields averaged over the period 1995-2019. Each color represents the spectrum of a different reanalysis product. Each scale value approximately corresponds to the spacing of the grid from which wavelet coefficients contributes to the energy at that scale.
    }
    \label{fig:wavelets}
\end{figure}

The results obtained from wavelets decomposition \ref{sec:methods_effRes} enabled the assessment of the effective resolution of the precipitation events at different spatial scales. The percentage energy spectra for the period 1995-2019 (Figure \ref{fig:wavelets}) represent the proportion of precipitation events occurring at different spatial scales during this period. The ERA5 spectrum indicates a different behaviour of this product compared to other reanalyses. Specifically, more than 90\% of precipitation events in ERA5 are reproduced at the meso-$\beta$ and upper scales (above 20 km). As expected, ERA5 consistently represents fewer events than higher-resolution models at meso-$\gamma$ and smaller scales (below 20 km). 
Additionally, significant differences are observed between convection-permitting and parametrized-convection regional models. While all regional products exhibit a higher component of fine-scale events (below 20 km) compared to ERA5, only convection-permitting reanalyses present a substantial fraction ($\sim 20\%$) of events at spatial scales below 10 km. Furthermore, the energy spectrum of these two categories displays distinct shapes. The convection-permitting reanalyses spectra result flatter than parametrized-convection models and report about the same percentage of events across meso-$\beta$ and  meso-$\gamma$ scales (from $\sim$ 10 km to $\sim$ 150 km). 
A single number representing the average effective resolution can be obtained by establishing a percentage energy threshold below which a product is considered unsatisfactory in reproducing precipitation patterns. However, the ranges of represented resolution discussed here may offer more significant insights than simply presenting effective resolution as a binary threshold distinguishing between skilled and unskilled scales. 

The spectra obtained need to be interpreted also in light of some possible influence of the interpolation methods on the results obtained.
Bilinear interpolation was employed to reconstruct precipitation fields at different scales (see Section \ref{sec:methods_effRes}). This method ensures a smooth transition between values at neighbouring grid points but may introduce features in precipitation fields at scales smaller than the original grid spacing of reanalyses. However, an alternative approach for downscaling to a finer grid, such as nearest-neighbor interpolation, introduces artifacts in the high-resolution field. These artifacts, such as sharp edges delineating transitions from one grid box to the next, affect the wavelet decomposition, resulting in increased wavelet coefficient values at the smallest scales to reconstruct these defined edges. For these reasons, bilinear interpolation was preferred.
The role of the gradual transition achieved through bilinear interpolation in downscaling affects the energy spectra by creating a smooth transition from spatial scales where wavelet coefficients significantly differ from zero in the original precipitation field to those scales where wavelet coefficients should be zero because they are smaller than the grid spacing. This transition does not substantially alter the characteristics of a spectrum and enables comparison of spectra across different reanalyses. This comparability is especially achievable because the grid spacing of reanalyses is accurately known, in contrast to their effective resolution. This knowledge aids in interpreting the spectrum, as it allows us to approximately identify the scales at which the impact of bilinear interpolation is expected (e.g. below 30 km for ERA5).

\begin{figure}[h!]
    \centering
    \includegraphics[width=1.1\textwidth]{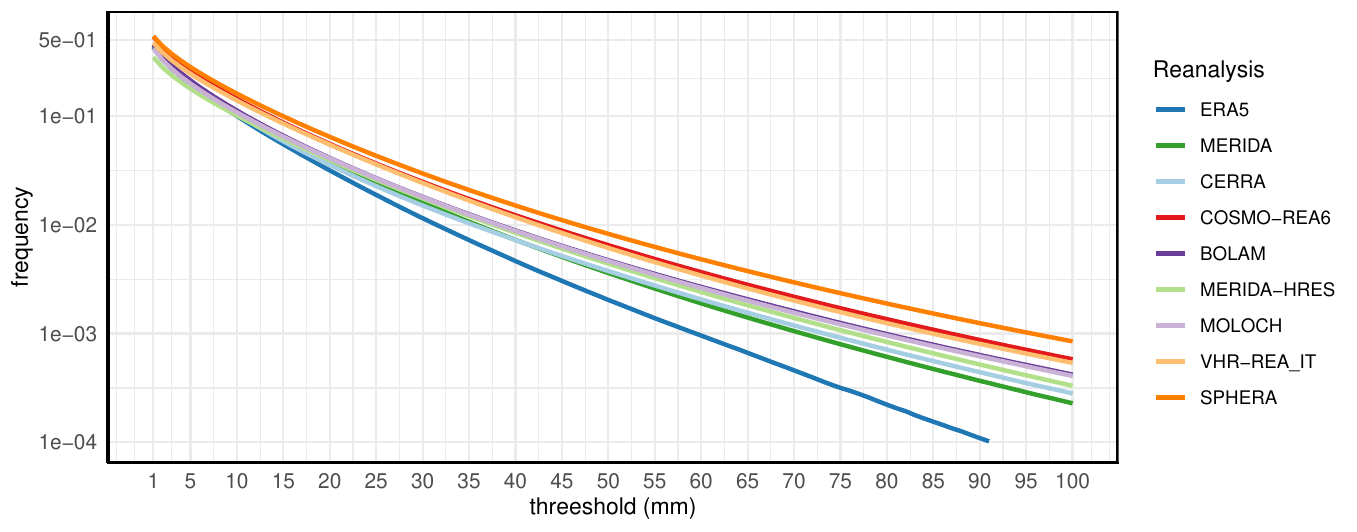}
    \caption{\textbf{Frequency Distribution of Daily Precipitation Intensities.} \\Frequency (y-axis, fraction of days, log-scale) at which daily precipitation amounts exceed a threshold (x-axis, mm, linear scale). Each color represents the distribution of a different reanalysis product.}
    \label{fig:freq}
\end{figure}

The frequency distributions of daily precipitation were also obtained for the same reference period 1995-2019 (Figure \ref{fig:freq}). 
Extremes of 1 mm and 100 mm were chosen such that the first point of the distribution (from the left) can be interpreted as the wet-day frequency at the Italian scale, while the last point on the right represents the frequency of days with precipitation exceeding 100 mm.

ERA5, with its coarse resolution, exhibits fewer occurrences of precipitation events exceeding 20 mm. This result supports the hypothesis that a coarser effective resolution leads to a smoother precipitation field compared to those reconstructed by models that have a finer effective resolution, with the effect of an average underestimation of intense precipitation.
As the precipitation threshold increases, the differences of the higher-resolution products from ERA5 increase.  

Higher-resolution products do not display a clear relationship between their effective resolution and the frequency of intense precipitation. Specifically, some of these products (like COSMO-REA6, VHR-REA\_IT, SPHERA) consistently exhibit slightly higher frequencies compared to other regional products across all threshold values. This suggests that these variations are more likely influenced by factors beyond resolution. Further insights into general wet or dry tendencies can be obtained by comparing reanalysis with observational averages (see Section \ref{sec:res_clim}).

\subsection{Validation against observations}\label{res:daily}

\subsubsection{Accuracy of daily precipitation}
\begin{figure}[h!]
    \centering
\includegraphics[width=1\textwidth,left]{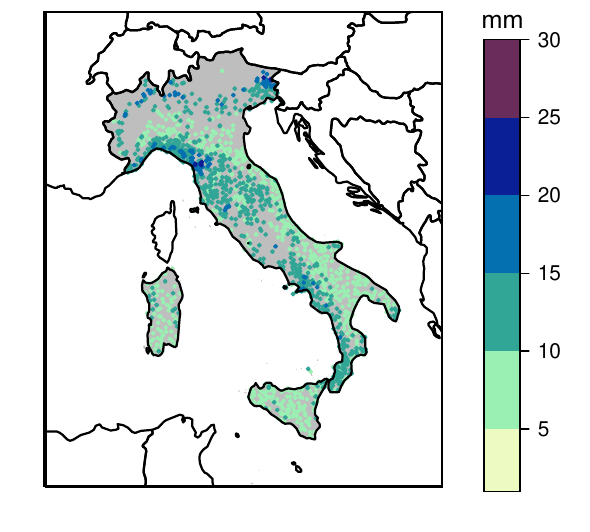}
    \caption{\textbf{Threshold Values for SEEPS Analysis.} \\Threshold values between 'light precipitation', and 'heavy precipitation' based on 2016-2020 weather station records.}
    \label{fig:thrs}
\end{figure}

The ability of each product to distinguish between 'dry', 'light precipitation', and 'heavy precipitation' against weather station measurements (Section \ref{sec:dewetra}) was evaluated using the SEEPS Skill Score (SS) as detailed in Section \ref{sec:met_seeps}. 
The thresholds delineating 'light precipitation' and 'heavy precipitation' were determined as outlined in Section \ref{sec:met_seeps} and vary from 10 mm to 25 mm due to different local climatologies over the Italian area (Figure \ref{fig:thrs}).

\begin{figure}[h!]
    \centering
    \includegraphics[width=1.1\textwidth]{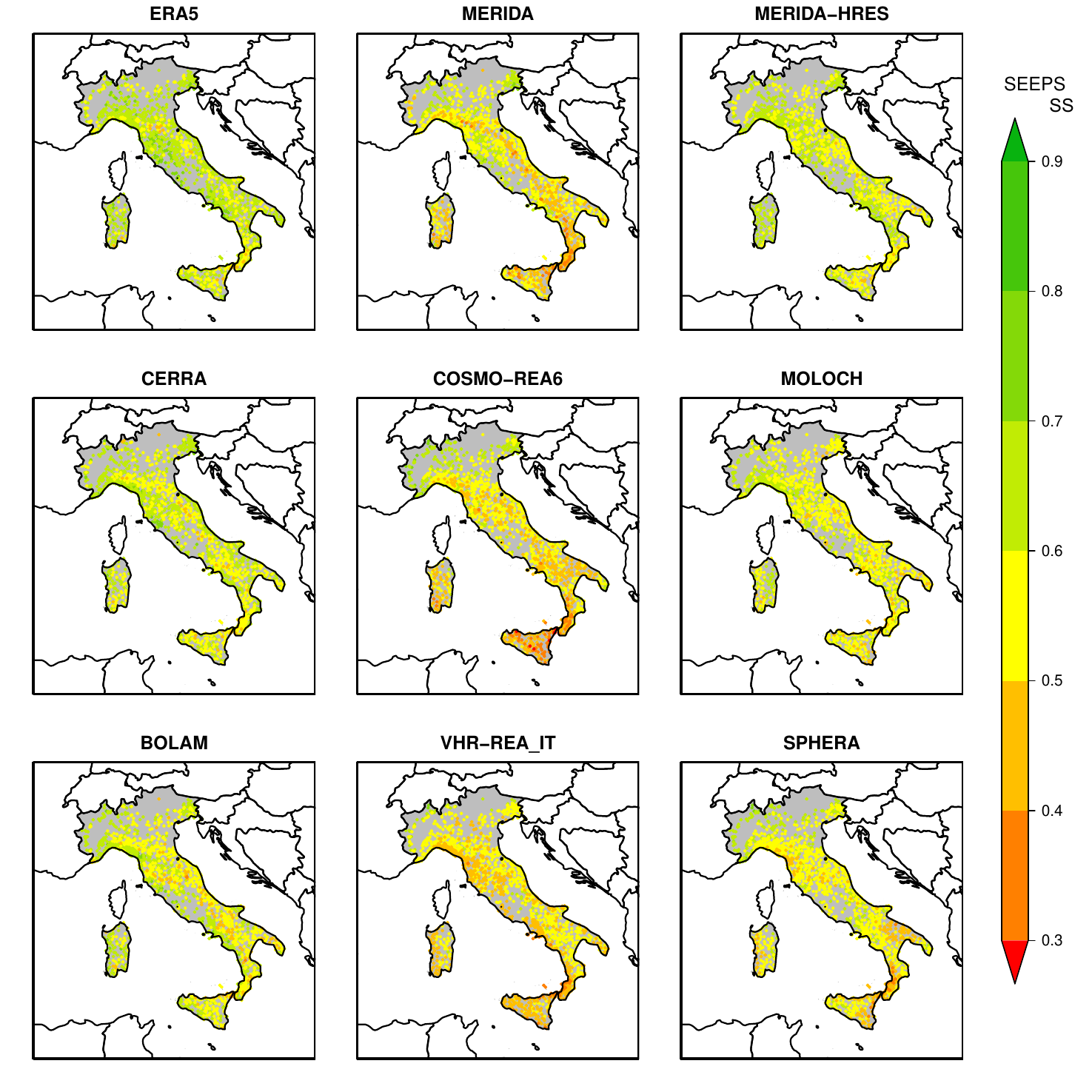}
    \caption{\textbf{SEEPS Skill Score Values.} \\SEEPS Skill Score values at each weather station position within the reference period 2016-2020 (ERA5, MERIDA, CERRA, MERIDA-HRES, VHR-REA\_IT) and 2016-2019 (COSMO-REA6, BOLAM, MOLOCH, SPHERA).}
    \label{fig:SEEPS}
\end{figure}

The maps depicting the SS values for each weather station, averaged over the reference periods, support the identification of error patterns across the Italian territory (Figure \ref{fig:SEEPS}). The low station density in the Alpine area can be attributed to observation scarcity in that area, but also to the exclusion of stations that do not meet the criteria outlined in Section \ref{sec:methods_verifobs}, particularly during winter to prevent undercatchment due to snow. Average values of SEEPS SS are presented in Table \ref{tab:seeps}, left part.
For the interpretation of the results, it is important to consider that the averages in time result from a different number of days across different seasons that match the significance criteria illustrated in Section \ref{sec:dewetra}.

ERA5 demonstrates notable performance across the entire Italian territory. Higher-resolution reanalyses, since representing precipitation events at a finer scale, result penalized in point-based scores. Indeed, smaller precipitation events are less likely to exactly match with observations at specific points. This issue is commonly known as the 'double-penalty' effect \citep{jermey_2016}. Nevertheless, CERRA and MERIDA-HRES report the highest SEEPS skill scores among regional and
convection-permitting reanalyses respectively, in line with those of ERA5, with a more detailed spatial description of the precipitation fields. MERIDA displays lower performance in terms of SEEPS over the Apennines, whereas VHR-REA\_IT and SPHERA show lower scores in southern Italy and along the north-western coast. MOLOCH and BOLAM present scores marginally lower than those of MERIDA-HRES and CERRA, albeit without a pattern linked to specific areas. COSMO-REA6 exhibits the lowest scores, with particularly low values in South Italy and Sicily island. 

The seasonal analysis highlights lower SEEPS SS during summer. This phenomenon is directly linked to the predominant convective nature of summer precipitation, which is characterized by smaller spatial scales and it is consequently more affected by model uncertainties, as detailed in Section \ref{sec:met_seeps}. At smaller scales, it is also more challenging to correctly place precipitation patterns, possibly leading to misplacement errors, resulting in low SEEPS SS values. To assess the contribution of misplacement, the comparison was repeated using a neighborhood area with a 15 km radius around each weather station. 

\begin{figure}[h!]
    \centering
    \includegraphics[width=1.1\textwidth]{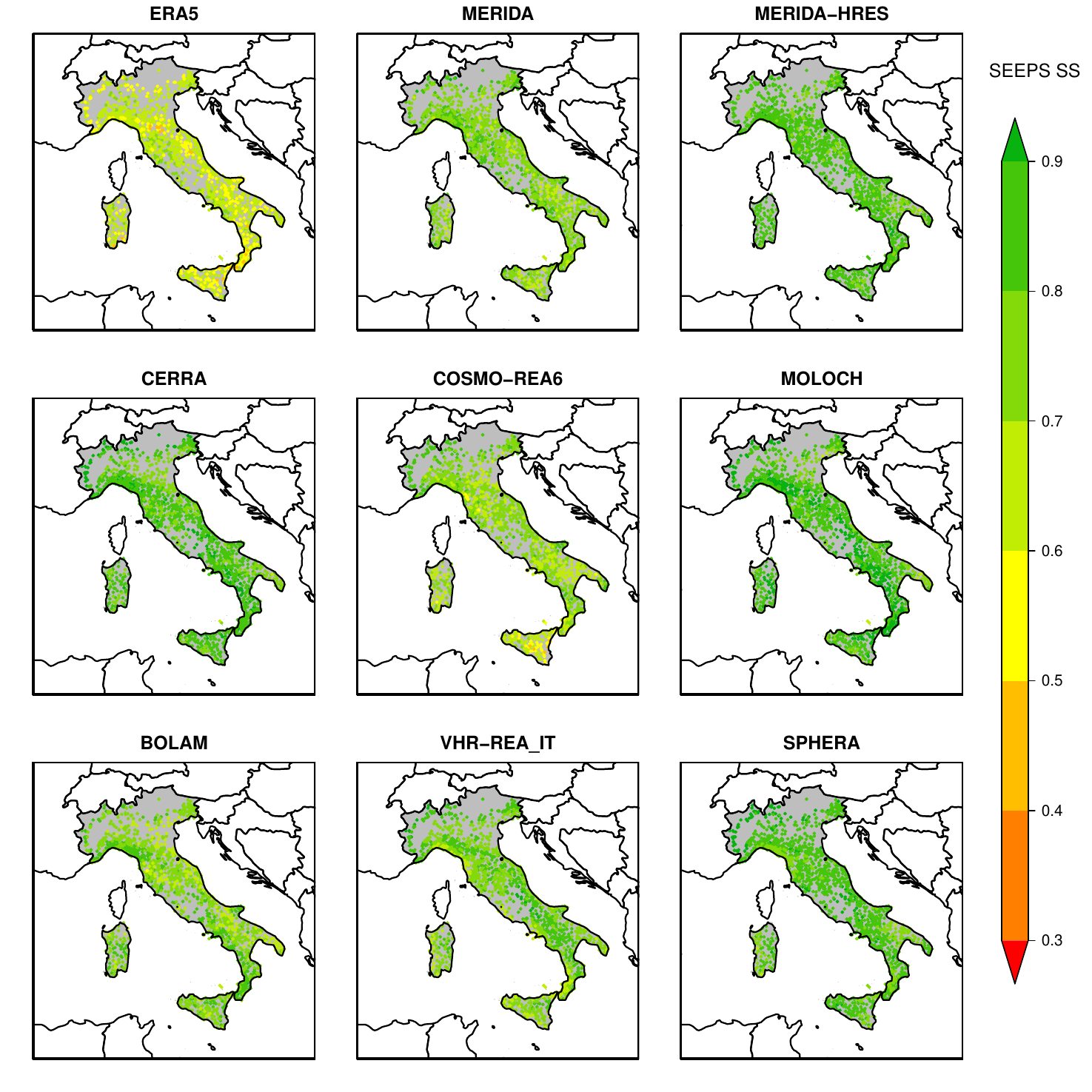}
    \caption{\textbf{Regionalized SEEPS Skill Score Values.} \\SEEPS Skill Score values obtained from the best reanalysis grid point value within 15 km from each weather station. Reference periods as in Figure \ref{fig:SEEPS}. }
    \label{fig:SEEPS_reg}
\end{figure}

The 15 km relaxation condition was selected to be representative of the lower boundary of the meso-$\gamma$ scale, which is the typical scale of daily precipitation \citep{thunis_scales_1996}. Moreover, it is roughly equivalent to half of the ERA5 grid spacing. Among all the selected reanalysis grid points, the SEEPS analysis was performed again using the reanalysis grid point which provide the precipitation value closest to that of the weather station. This approach allows high-resolution models to accommodate misplacement errors of up to approximately 15 km. Under these conditions, most error patterns disappear and higher-resolution products give evidence of significantly higher scores than ERA5 ones (see Figure \ref{fig:SEEPS_reg} and the right part of Table \ref{tab:seeps}). 
The ERA5 scores remain unchanged because the 15 km radius still allowed the selection of the same nearest neighbor grid point as before. The increase in the SEEPS Skill Scores indicates that reanalyses can often correctly identify the local precipitation category (among 'dry', 'light', and 'heavy'), though not at the weather station position, but within a 15 km surrounding area. This uncertainty arises from the inherently chaotic nature of the atmosphere, which imposes increasingly greater limitations on predictability as the scale becomes smaller (Section \ref{sec:predictability}). Higher-resolution reanalyses are particularly affected by these kinds of limitations, as they aim to resolve more challenging processes at finer scales. This is evidenced by the significant improvement in their SEEPS Skill Scores under the 15-km relaxation condition (Table \ref{tab:seeps}).

\subsubsection{Deviations from observational climatology}\label{sec:res_clim}

The annual cumulative precipitation averaged from 1995 to 2019 at the native resolution of each reanalysis highlights the higher level of detail provided by higher-resolution products, particularly in areas of complex orography (Figure \ref{fig:climmaps}). CERRA shows some unique issues, discussed in the supplementary material.

The relative bias (as per equation \ref{eq:percbias} in Section \ref{sec:met_gridd}) of reanalysis climatological fields compared to the corresponding annual observational data reveals similarities but also some strong differences among the different products (Figure \ref{fig:climpercbias}). 
The analysis underscores a notable tendency towards wet biases in northern Italy and dry biases in central and southern Italy.
Specifically, CERRA and VHR\_REA-IT exhibit a consistent wet bias throughout the Alps. ERA5 shows wet biases in the western Alps and the Po Valley. MERIDA and MERIDA-HRES demonstrate a comparable pattern, featuring more localized wet biases primarily in the Po Valley and Central Alps. However, MERIDA shows dry biases across the southern Apennines and the islands, a feature not present in MERIDA-HRES. BOLAM and MOLOCH display similar spatial patterns characterized by sporadic wet and dry biases, with MOLOCH generally exhibiting closer alignment with observational data.
COSMO-REA6 and SPHERA exhibit the lowest biases over the Alps, which becomes more pronounced in the western region. Conversely, SPHERA demonstrates a wet bias across the Po Plain and southern Italy, whereas COSMO-REA6 reveals moderate dry biases along the Apennines and a strong dry bias over Sicily and Sardinia. Similarly, VHR\_REA-IT indicates a dry bias along the western coasts of the Italian peninsula and the islands.

To check if some of the small values of relative bias (between -10\% and 10\%, e.g., in the Po Plain for VHR-REA\_IT or in the eastern Alps for SPHERA) are a result of the superposition of monthly biases of opposite sign (sometimes dry and sometimes wet), annual averages of monthly dry and wet biases were also computed separately (figures not shown).
It was verified that their patterns mostly do not overlap, confirming the average good performance of reanalyses where the relative bias is small.

A complementary aspect to consider alongside the spatial distribution of climatological biases is the bias variation across different months of the year. This question was addressed by obtaining the spatial mean of the monthly precipitation average for the period 1995-2019 for all reanalysis products and the observational gridded dataset (Figure \ref{fig:climmonthlycycle}). It is known that the precipitation seasonality is different in various regions of the Italian peninsula (as shown in Figure 14 of \cite{crespi_2018}). However, the spatial average over the Italian territory  (Figure \ref{fig:climmonthlycycle}) shows that reanalyses generally tend to overestimate precipitation during spring and summer (except MOLOCH and COSMO-REA6), while only certain products display underestimation during autumn (VHR\_REA-IT, SPHERA, MERIDA, COSMO-REA6).

To provide a variability range for those averages, the relative wet and dry monthly biases were calculated and then separately spatially averaged (Figure \ref{fig:dry_wet_avgs}). Based on this analysis, it is found that all reanalyses share an average dry relative bias of approximately $-20\%$ across all months. Moreover, in wet bias areas, the relative bias is, on average, $+20\%$, but only MOLOCH and COSMO-REA6 maintain this value approximately constant throughout the year. Other products exhibit larger wet relative biases from April to September, reaching up to $+60\%$ for MERIDA in July.

\subsubsection{Temporal consistency of annual precipitation totals}\label{sec:res_trend}
Finally, the comparison between annual reanalysis rainfall totals and observational data provides insight into the comprehension of decadal precipitation trends and their attribution (Figure \ref{fig:trend}). An actual climate signal regarding precipitation is expected to be captured in both observational data and reanalyses. Consequently, the series of annual differences should not show discernible long-term trends. Surprisingly, nearly all products exhibit a significant trend in the series of annual differences against observations. Therefore, the precipitation trends observed in reanalyses over Italy must be partially attributed to a divergence from observations. The magnitude of this disparity can be assessed by dividing the value of the trend (expressed in mm/decade) by the typical annual observational rainfall amount (Table \ref{tab:trend}). While ERA5, MERIDA, and CERRA exhibit only a modest ($3\%$ per decade) yet still significant deviation, MERIDA-HRES exhibits a slightly higher trend ($4\%$ per decade). COSMO-REA6, BOLAM and MOLOCH present a marked trend ($7\%$ per decade) while
SPHERA demonstrates the highest increase, exceeding $10\%$ per decade. Conversely, VHR-REA\_IT shows the smallest ($2\%$ per decade) trend, which is statistically less significant.
It should be noted that different areas of the Italian territory exhibit this behaviour with different magnitudes, as shown in the Supplementary material (Figure S4). These deviations should be considered when using reanalyses to compute precipitation trends over Italy.

\section{Conclusions}\label{sec:conclusions}

An inter-comparison and validation of precipitation fields from nine reanalysis products was conducted, providing insights into the capabilities of the different products in reproducing precipitation fields and their accuracy across different spatial scales. Initially, a wavelet decomposition was applied to daily precipitation fields, to investigate the effective resolution of the different reanalysis products. This approach allowed a characterization of the precipitation energy spectra across spatial scales, highlighting the strong differences among global (ERA5) and regional reanalyses, and convection-permitting vs. parametrized models. This analysis indicated that only convection-permitting products are able to represent a substantial fraction of precipitation events below 10 km, while regional models present an energy shift towards finer scales than ERA5 (10 - 30 km). Furthermore, the frequency analysis demonstrated that ERA5 consistently represents fewer rainy days exceeding the 20 mm or higher thresholds than higher-resolution products.

However, the representation of small-scale precipitation events exposes higher-resolution models to misplacement issues when compared to observational data. This phenomenon was quantified by assessing reanalysis daily precipitation fields against weather station data. The findings revealed lower skills during summer, consistent with the challenges represented by small-scale convection processes during the summer months. Some products (CERRA, MERIDA-HRES) exhibited good performance comparable to the coarse-resolution ERA5, indicating a minor impact of misplacement. Other products showed lower skills, often influenced by geophysical elements such as mountains and coastlines. Notably, a 15-km neighboring criterion in the calculation of the score improved the performances of regional models and highlighted the significant role of misplacement.

In addition to misplacement, climatological comparisons revealed wet biases during spring and summer in Northern Italy and dry biases during autumn and winter in Southern Italy, albeit with some differences across the different products. Finally, the temporal stability of annual precipitation trends was assessed by averaging accumulated annual precipitation across the entire Italian territory. Most reanalyses show a significant trend in annual differences with observations. This analysis indicates that deviations from observations can explain at least part of the precipitation trend signals reproduced by reanalyses over Italy.

In conclusion, this study has examined the capabilities of global and regional reanalyses, including both parameterized-convection and convection-permitting models, in representing precipitation patterns and trends across Italy. It highlighted the strengths and limitations of these products, enabling potential users to employ them more effectively for research and practical applications. 
This understanding will aid in making informed decisions on climate-related issues in Italy and beyond.

\clearpage
\section*{Funding Information}
The PhD of the co-author Francesco Cavalleri was activated pursuant to DM 352 and is co-sponsored by PNRR funds and R.S.E. s.p.a. The PNRR funds come from the EU Next-generation programme. This work has been financed by the Research Fund for the Italian Electrical System under the Three-Year Research Plan 2022–2024 (DM MITE n. 337, 15.09.2022) in compliance with the Decree of 16 April 2018. This work has been financed by Research Funds from the Italian Ministry for University and Research (PRIN 2022 - CN4RWK – CCHP-ALPS – Climate Change and HydroPower in the Alps, funded by the European Union (Programme Next Generation EU)).
Veronica Manara was supported by the “Ministero dell'Università e della Ricerca” of Italy [grant FSE – REACT EU, DM 10/08/2021 n. 1062]. 

\section*{Declaration of interests}
The authors declare that they have no known competing financial interests or personal relationships that could have appeared to influence the work reported in this paper.

\section*{Data Availability Statement}

ERA5 and CERRA data are openly available from Copernicus Climate Change Service (C3S) in the Climate Data Store (CDS), respectively: ``ERA5 hourly data on single levels from 1940 to present'' at \url{http://doi.org/10.24381/cds.adbb2d47} (Accessed on 24-JAN-2024) and 
``CERRA sub-daily regional reanalysis data for Europe on single levels from 1984 to present'' at \url{http://doi.org/10.24381/cds.622a565a} (Accessed on 6-DEC-2023).
\par COSMO-REA6 data are openly available from the DWD opendata-FTP server: ``daily/2D/TOT\_PRECIP'' at \\\url{https://opendata.dwd.de/climate_environment/REA/COSMO_REA6/}
\par MERIDA and MERIDA-HRES data are openly available from the RSE repository: ``PREC'' at \url{https://merida.rse-web.it/}
\par VHR-REA\_IT data are openly available from CMCC Data Delivery System: ``Total precipitation amount'' at \\ \url{https://dds.cmcc.it/#/dataset/era5-downscaled-over-italy}
\par BOLAM and MOLOCH data are available upon reasonable request from the co-author Valerio Capecchi.
\par SPHERA total precipitation and 2m-temperature data are freely available on Zenodo repository (for total precipitation: \url{https://zenodo.org/records/10441408}). Other data outputs can be made available upon request from ARPAE data archive; contact person: co-author Davide Cesari.
\par UniMi/ISAC-CNR observational data supporting the findings of this study are available from the co-author Michele Brunetti upon reasonable request.
\par Regional ARPA weather station data have been made available for the research within the Resilience and Security Project of the Italian electrical system with the support of Marilena Barbaro (Ministry of Economic Development) and Carlo Cacciamani (Central Functional Centre, Department of Civil Defence), and may be available by directly contacting them. 


\clearpage
During the preparation of this work, the authors used ChatGPT in order to enhance readability. After using this tool, the authors reviewed and edited the content as needed and take full responsibility for the content of the publication.

\bibliographystyle{elsarticle-harv} 
\bibliography{cas-refs}
\clearpage

\section*{Tables}
\begin{table}[h!]
\caption{\normalsize \raggedright Details of the reanalysis products considered.}
\resizebox{0.49\paperheight}{!}{
\begin{tabular}{
  p{\dimexpr.15\linewidth-2\tabcolsep-1.3333\arrayrulewidth}
  |p{\dimexpr.15\linewidth-2\tabcolsep-1.3333\arrayrulewidth}
 p{\dimexpr.17\linewidth-2\tabcolsep-1.3333\arrayrulewidth}
  p{\dimexpr.15\linewidth-2\tabcolsep-1.3333\arrayrulewidth}
 p{\dimexpr.125\linewidth-2\tabcolsep-1.3333\arrayrulewidth}
  p{\dimexpr.3\linewidth-2\tabcolsep-1.3333\arrayrulewidth}
 p{\dimexpr.125\linewidth-2\tabcolsep-1.3333\arrayrulewidth} 
  }
\textbf{Name} & \textbf{Devel. group} & \textbf{NWP Model} & \textbf{I.C. and B.C.} & \textbf{Horiz. grid} & \textbf{Assimilated data} & \textbf{Conv-perm.} \\
\hline
ERA5 & ECMWF & IFS CY41R2 & / & 0.25° & see ERA5 User Guide & No \\ \hline
CERRA & ECMWF & ALADIN & ERA5 & 5.5 km & post-processed GNSS-RO (radio occultation) and GNSS-ZTD (zenith total delay) & No \\ \hline
COSMO-REA6 & DWD & COSMO & ERA-Interim & 6 km & continuous nudging of radiosondes, aircraft measurements, wind profiler and station data & No \\ \hline
MERIDA & RSE & WRF-ARW v3.9 & ERA5 & 0.07° & spectral nudging of $\Phi$, $u$ and $v$ wind components, and $T$. T2m observation nudging. & No \\ \hline
BOLAM & LaMMA & BOLAM & ERA5 & 0.07° & None & No \\ \hline
MERIDA-HRES & RSE & WRF & ERA5 & 0.04° & spectral nudging of $\Phi$, $u$ and $v$ wind components, and $T$. T2m observation nudging. Assimilation of MERRA2 aerosol data. & Yes \\ \hline
MOLOCH & LaMMA & MOLOCH & BOLAM & 0.0247° & None & Yes \\ \hline
VHR- -REA\_IT & CMCC & COSMO-CLM v5.0 & ERA5 & 0.02° & None & Yes \\ \hline
SPHERA & ARPAE & COSMO v5.05 & ERA5 & 0.02° & continuous nudging of $p$, $u$ and $v$ wind components, $hum.$ and $T$. & Yes \\
\hline
\end{tabular}}
\vspace{0.1cm}

\label{tab:nwp_comparison}
{\raggedright \footnotesize ECMWF: European Centre for Medium-range Weather Forecasts, DWD: Deutscher WetterDienst, RSE: Ricerca sul Sistema Energetico, LaMMA: Laboratorio di Monitoraggio e Modellistica Ambientale, CMCC: Centro euro-Mediterraneo sui Cambiamenti Climatici, ARPAE: Agenzia Regionale Prevenzione, Ambiente ed Energia dell'Emilia-romagna, MERRA2: Modern-Era Retrospective analysis for Research and Applications, version 2. \par}
\end{table}

\normalsize
\begin{table*}[h!]
\caption{\normalsize SEEPS skill score averaged over all stations. The reference period is 2016-2019 for COSMO-REA6, BOLAM, MOLOCH and SPHERA, and 2016-2020 for the other products. These spatial averages result from an uneven distribution of weather stations; hence, caution must be exercised when interpreting them as representative of the Italian average.}
\resizebox{0.7\paperwidth}{!}{
\begin{tabular}{lccccc|ccccc}
 & \multicolumn{5}{c}{\textbf{nearest grid point}} & \multicolumn{5}{c}{\textbf{best grid point within 15 km}} 
\\ \cline{2-11}
\multicolumn{1}{l|}{\textbf{Reanalisys}} & \multicolumn{1}{l}{DJF}  & \multicolumn{1}{l}{MAM}  & \multicolumn{1}{l}{JJA}  & \multicolumn{1}{l}{SON}  & \multicolumn{1}{l|}{year} & \multicolumn{1}{l}{DJF} & \multicolumn{1}{l}{MAM} & \multicolumn{1}{l}{JJA} & \multicolumn{1}{l}{SON} & \multicolumn{1}{l}{year} \\
\hline 
 \multicolumn{1}{l|}{ERA5} & \multicolumn{1}{l}{0.67} & \multicolumn{1}{l}{0.65} & \multicolumn{1}{l}{0.49} & \multicolumn{1}{l}{0.65} & 0.64 
& \multicolumn{1}{l}{0.67} & \multicolumn{1}{l}{0.65} & \multicolumn{1}{l}{0.49} & \multicolumn{1}{l}{0.65} & 0.64 \\ \hline  
\multicolumn{1}{l|}{MERIDA} & \multicolumn{1}{l}{0.59} & \multicolumn{1}{l}{0.58} & \multicolumn{1}{l}{0.41} & \multicolumn{1}{l}{0.55} & 0.55 & \multicolumn{1}{l}{0.78}  & \multicolumn{1}{l}{0.81} & \multicolumn{1}{l}{0.73} & \multicolumn{1}{l}{0.77} & 0.77 \\  \hline  
\multicolumn{1}{l|}{CERRA} & \multicolumn{1}{l}{0.67} & \multicolumn{1}{l}{0.64} & \multicolumn{1}{l}{0.46} & \multicolumn{1}{l}{0.62} & 0.61  & \multicolumn{1}{l}{0.87} & \multicolumn{1}{l}{0.87} & \multicolumn{1}{l}{0.83} & \multicolumn{1}{l}{0.85} & 0.85 \\ \hline  
\multicolumn{1}{l|}{COSMO-REA6} & \multicolumn{1}{l}{0.58} & \multicolumn{1}{l}{0.58} & \multicolumn{1}{l}{0.42} & \multicolumn{1}{l}{0.55} & 0.53 & \multicolumn{1}{l}{0.75} & \multicolumn{1}{l}{0.78} & \multicolumn{1}{l}{0.68} & \multicolumn{1}{l}{0.74} & 0.73  \\ \hline  
\multicolumn{1}{l|}{BOLAM} & \multicolumn{1}{l}{0.63} & \multicolumn{1}{l}{0.60} & \multicolumn{1}{l}{0.45} & \multicolumn{1}{l}{0.60} & 0.59 & \multicolumn{1}{l}{0.80} & \multicolumn{1}{l}{0.78} & \multicolumn{1}{l}{0.68} & \multicolumn{1}{l}{0.77} & 0.77  \\  \hline  
\multicolumn{1}{l|}{MERIDA-HRES} & \multicolumn{1}{l}{0.69} & \multicolumn{1}{l}{0.62} & \multicolumn{1}{l}{0.43} & \multicolumn{1}{l}{0.63} & 0.61  & \multicolumn{1}{l}{0.87} & \multicolumn{1}{l}{0.85} & \multicolumn{1}{l}{0.77} & \multicolumn{1}{l}{0.85} & 0.84 \\ \hline  
\multicolumn{1}{l|}{MOLOCH} & \multicolumn{1}{l}{0.64} & \multicolumn{1}{l}{0.58} & \multicolumn{1}{l}{0.41} & \multicolumn{1}{l}{0.61} & 0.58  & \multicolumn{1}{l}{0.89} & \multicolumn{1}{l}{0.88} & \multicolumn{1}{l}{0.83} & \multicolumn{1}{l}{0.88} & 0.87  \\  \hline  
\multicolumn{1}{l|}{VHR-REA\_IT} & \multicolumn{1}{l}{0.57} & \multicolumn{1}{l}{0.54} & \multicolumn{1}{l}{0.42} & \multicolumn{1}{l}{0.54} & 0.51  & \multicolumn{1}{l}{0.79} & \multicolumn{1}{l}{0.81} & \multicolumn{1}{l}{0.80} & \multicolumn{1}{l}{0.79} & 0.78 \\ \hline  
\multicolumn{1}{l|}{SPHERA} & \multicolumn{1}{l}{0.57} & \multicolumn{1}{l}{0.59} & \multicolumn{1}{l}{0.47} & \multicolumn{1}{l}{0.58} & 0.55  & \multicolumn{1}{l}{0.08} & \multicolumn{1}{l}{0.87} & \multicolumn{1}{l}{0.86} & \multicolumn{1}{l}{0.85} & 0.84 \\ \hline  
\end{tabular}}
\label{tab:seeps}
\end{table*}

\begin{table*}[h!]
\caption{\normalsize Decadal trends of reanalysis minus observational annual total precipitation averages over Italy. Trends were obtained using the Theil-Sen method \citep{sen_1968}. Percentual values in the third column were normalized with the 1995-2019 observational climatological value. Significance was calculated using the Mann-Kendall method.}
\resizebox{0.7\paperwidth}{!}{
\begin{tabular}{l|ccc}

\textbf{Reanalysis} & \textbf{(rean - obs) [mm/10yr]}	& \textbf{(rean - obs)/obs [\%/10yr]} & \textbf{p-value (M-K)} \\ \hline  
ERA5	   &31	&3\%	& $<$0.01\\ \hline  
MERIDA	   &31	&3\%	&0.01\\ \hline  
CERRA	   &36	&3\%	&$<$0.01\\ \hline  
COSMO-REA6 &68	&7\%	&$<$0.01\\ \hline  
BOLAM	   &69	&7\%	&$<$0.01\\ \hline  
MERIDA-HRES&33	&4\%	&$<$0.01\\ \hline  
MOLOCH	   &75	&7\%	&$<$0.01\\ \hline  
VHR-REA\_IT &23	&2\%	&0.05\\  \hline  
SPHERA	   &98	&10\%	&$<$0.01\\ \hline 
\end{tabular}}
\label{tab:trend}
\end{table*}

\clearpage
\section*{Figures}

\begin{figure}[h!]
    \centering
    \includegraphics[width=1.1\textwidth]{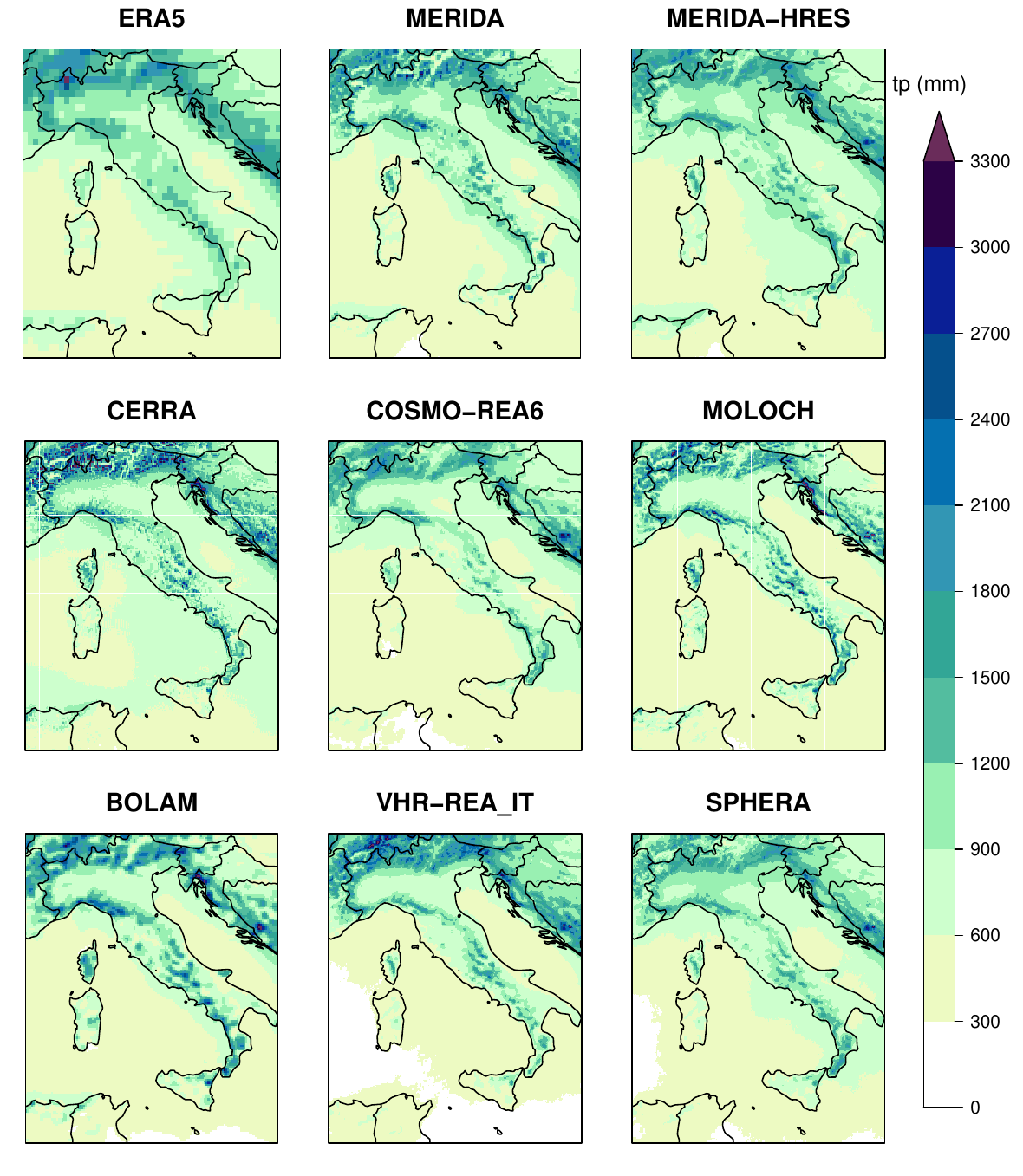}
    \caption{\textbf{Annual Precipitation Climatology Maps.}  \\The annual cumulative precipitation averaged 1995-2019  at the native resolution of each reanalysis.}
    \label{fig:climmaps}
\end{figure}

\begin{figure}[h!]
    \centering
    \includegraphics[width=1.1\textwidth]{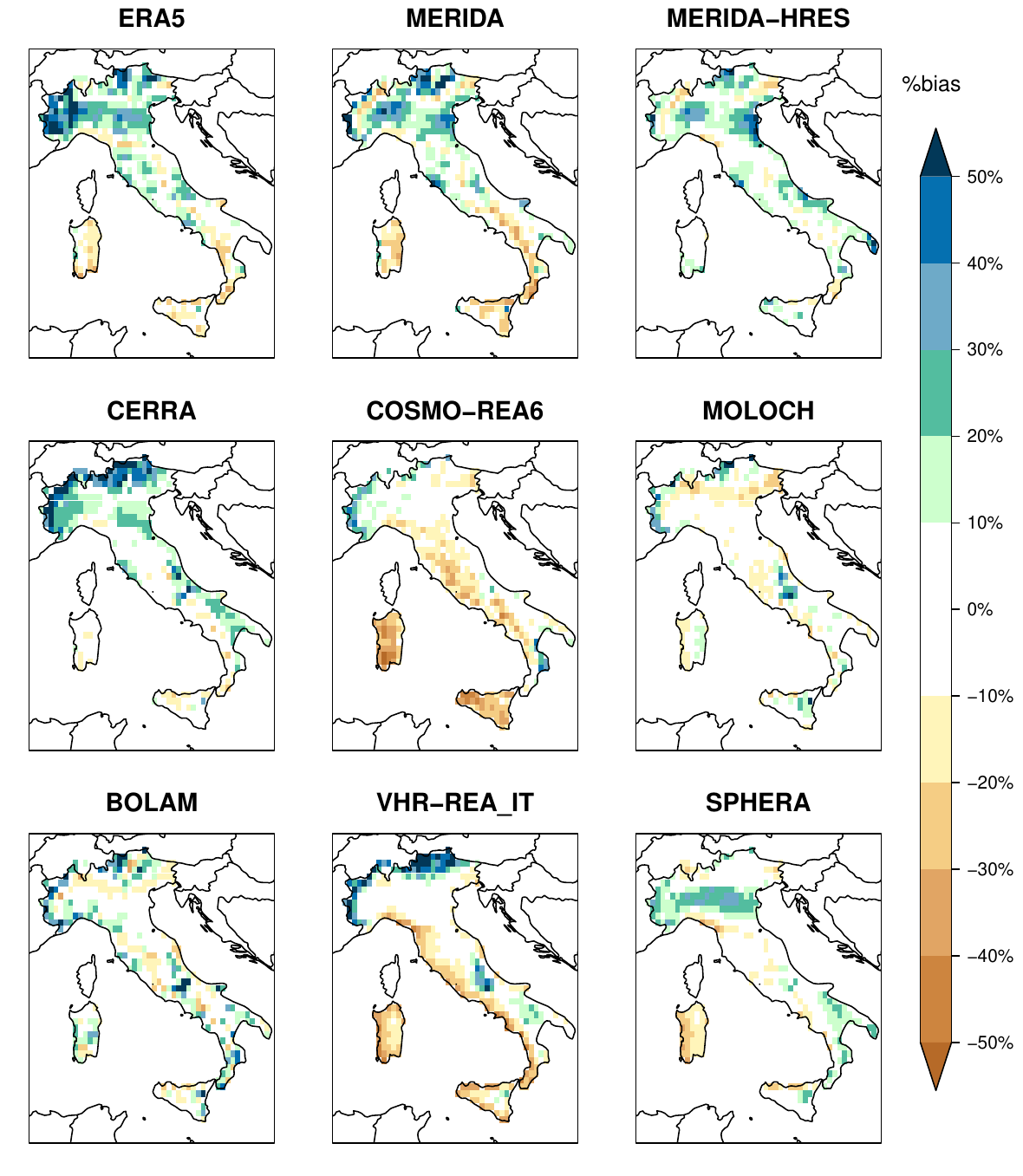}
    \caption{\textbf{Climatological Relative Bias Maps.} \\The percentual bias against observations of the annual 1995-2019 climatological precipitation fields. Each reanalysis was evaluated after conservative upscaling at the common ERA5 resolution.}
    \label{fig:climpercbias}
\end{figure}

\begin{figure}[h!]
    \centering
    \includegraphics[width=1.1\textwidth]{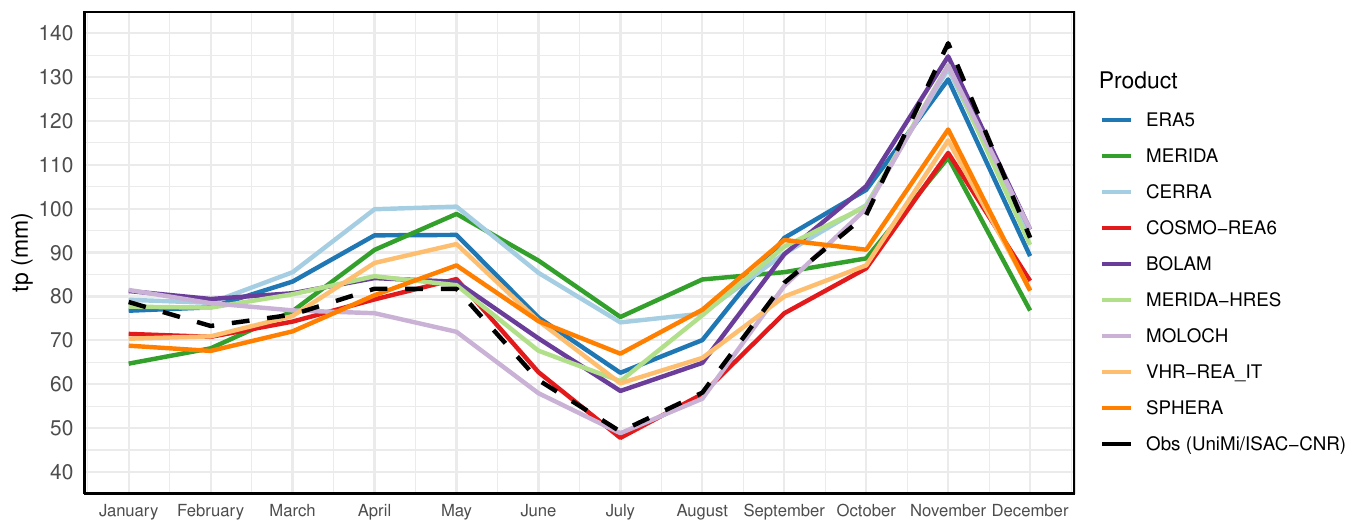}
    \caption{\textbf{Average Monthly Precipitation Climatology Graph.} \\The 1995-2019 climatological monthly cumulate precipitations averaged over the Italian territory. Each color represents a different reanalysis product. Reference observational climatology is shown in black.}
    \label{fig:climmonthlycycle}
\end{figure}

\begin{figure}[h!]
    \centering
    \includegraphics[width=1.05\textwidth]{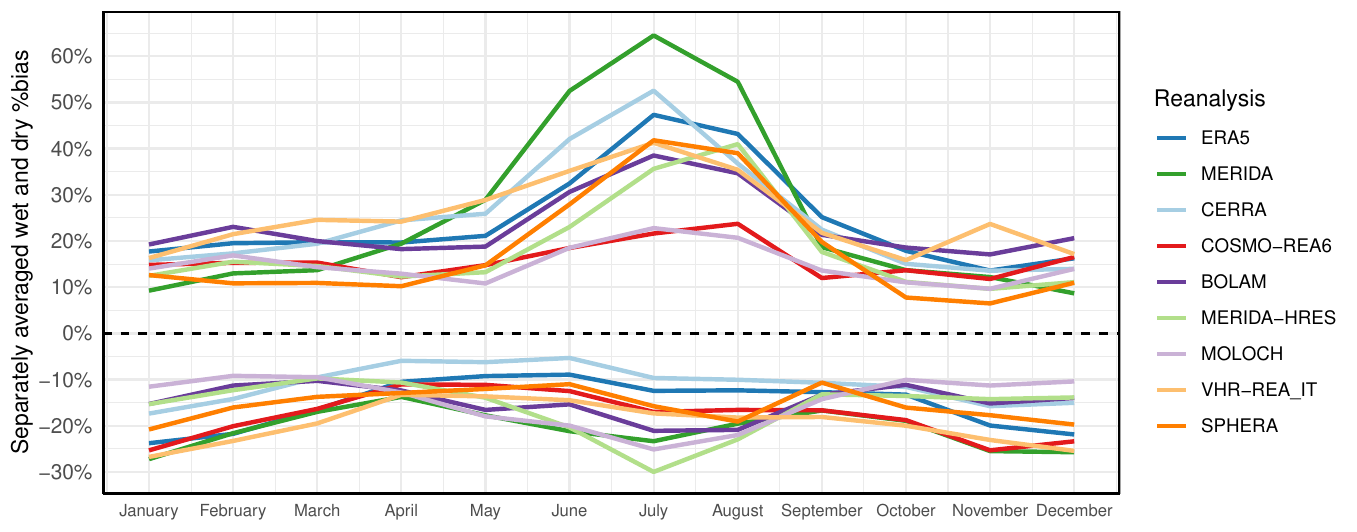}
    \caption{\textbf{Dry and Wet Average Monthly Precipitation Biases against Observations.} \\The 1995-2019 climatological monthly precipitation relative bias averaged over the Italian territory. Above, only grid points with a wet bias were averaged. Below, only the ones with a dry bias. Each color represents a different reanalysis product.}
    \label{fig:dry_wet_avgs}
\end{figure}

\begin{figure}[h!]
    \centering
    \includegraphics[width=1.1\textwidth]{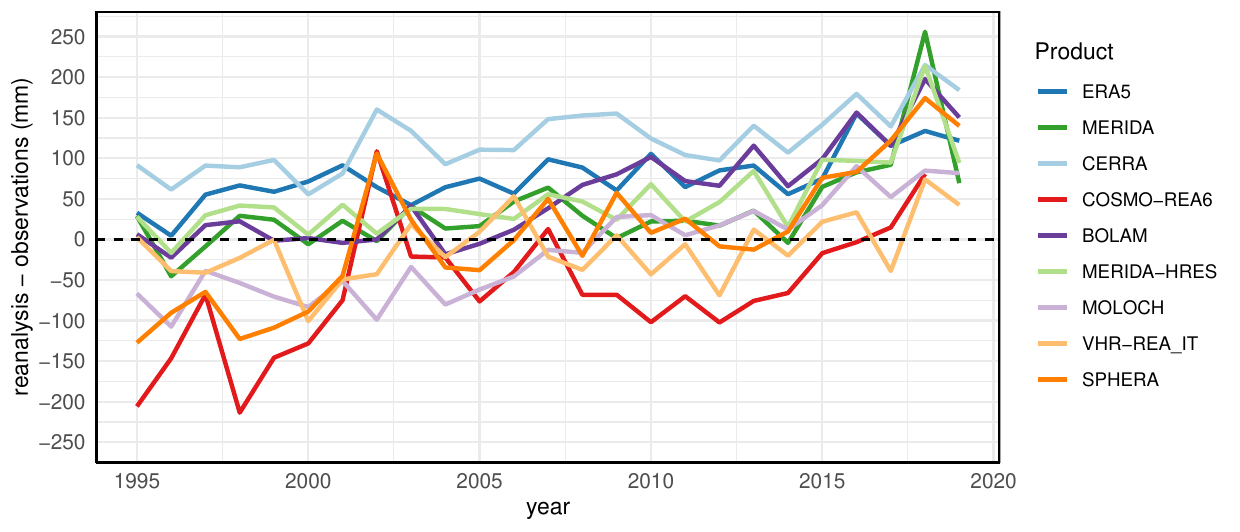}
    \caption{\textbf{Graph of the Annual Total Precipitation Deviation from Observation.} \\The annual differences between Italian average annual precipitation from reanalysis and observations (UniMi/ISAC-CNR). Each color represents the deviation of a different reanalysis product.}
    \label{fig:trend}
\end{figure}









\end{document}